\documentclass[a4paper, fleqn, usenatbib]{mnras} 
\pdfoutput=1
\usepackage{mathptmx}

\usepackage{graphicx}	
\usepackage{amsmath}	
\usepackage{amssymb}	
\usepackage{color}

\title[Lagrangian Methods Of Cosmic Web Classification]{Lagrangian Methods Of Cosmic Web Classification}

\author[J. D. Fisher, A. Faltenbacher and M.S.T. Johnson]{J. D. Fisher\thanks{E-mail: justindavidfisher@gmail.com}, A. Faltenbacher and M.S.T. Johnson\\
Department of Physics, University of Witwatersrand, Braamfontein, 2000, South Africa}	

\date{Accepted XXX. Received YYY; in original form ZZZ}

\pubyear{2015}

\begin{document}
\label{firstpage}
\pagerange{\pageref{firstpage}--\pageref{lastpage}}
\maketitle

\begin{abstract}
  The cosmic web defines the large scale distribution of matter we see in the Universe today. Classifying the cosmic web into voids, sheets, filaments and nodes allows one to explore structure formation and the role environmental factors have on halo and galaxy properties. While existing studies of cosmic web classification concentrate on grid based methods, this work explores a Lagrangian approach where the V-web algorithm proposed by \cite{Hoffman2012} is implemented with techniques borrowed from smoothed particle hydrodynamics. The Lagrangian approach allows one to classify individual objects (e.g. particles or halos) based on properties of their nearest neighbours in an adaptive manner. It can be applied directly to a halo sample which dramatically reduces computational cost and potentially allows an application of this classification scheme to observed galaxy samples. Finally, the Lagrangian nature admits a straight forward inclusion of the Hubble flow negating the necessity of a visually defined 
threshold value which is commonly employed by grid based classification methods.
\end{abstract}



%

\section{Introduction}

The large scale distribution of matter throughout the Universe is originally described by \cite{Bond1995} as the cosmic web, exhibiting a filamentary structure with regions of increasing density classified as voids, sheets, filaments and nodes. Cosmological numerical simulations are used to explore the large scale structure and have contributed significantly to the establishment of the $\Lambda$CDM model of structure formation. Cold dark matter (DM) is said to constitute $\sim85\%$ of the gravitating mass within the Universe, thus understanding the kinematics of DM allows one to probe into the evolution of the Universe in both the linear and non-linear regime.  We use pure DM simulations, neglecting the baryonic matter component, to represent the N-body problem as a fluid under gravitational evolution using the Vlasov - Poisson equations \citep{kuhlen2012}.  

Galaxy morphology has been shown to vary with the cosmic environment \citep{Dressler1980, Blanton2003, Gao2005b, Faltenbacher2009} and exploring this link is crucial to the development of a consistent cosmological theory of galaxy formation and evolution. Galaxy redshift surveys and DM N-body simulations have shown a significant statistical correlation in filamentary structure with the different web types thought to be a result of small density perturbations in the early universe that have undergone gravitational collapse.  

Initial attempts to classify different web types with kinematical information began with \cite{Hahn2007a,Aragon-Calvo2007a} and \cite{Forero-Romero2008} using Tidal Torque Theory, evaluating the Hessian of the gravitational potential. Since then, more sophisticated algorithms have been introduced to describe the topology of the phase space and identify substructure. \cite{Aragon-Calvo2010a} used the SpineWeb framework based on the watershed algorithm \citep{Platen2007} to segment the density field. \cite{Falck2012} showed that one is able to classify morphological structures using the ORIGAMI method: counting the number of orthogonal folds in the Lagrangian phase space sheet. Various methods have borrowed techniques from discrete Morse theory \citep{Colombi2000} using the Delaunay Tessellation Field Estimator \citep{Schaap:2000se} to identify substructure in the the cosmological density fields \citep{Aragon-Calvo2007a, Sousbie2011a, Aragon-Calvo2010a}.

\cite{Hoffman2012} introduced a classification scheme based on the shear of the velocity field (V-web) and showed it is able to improve on the resolution of the tidal tensor based method (T-web), resolving smaller structures which allows one to better study halo properties. DM simulations are characterised by the density and velocity fields with these two fields being related by Eq. ~\ref{conservation}, the familiar mass conservation equation\footnote{ $\frac{D\rho}{Dt} = \frac{\partial \rho}{\partial t} + \vec{v} \cdotp \nabla\rho$ is the comoving derivative}:
\begin{equation}
  \frac{1}{\rho}\frac{D\rho}{Dt} = -(\nabla \cdotp \vec{v})\ . 
  \label{conservation}
\end{equation}
\cite{Hoffman2012} provides evidence of the velocity divergence field tracing the density field; a divergent velocity field being associated with under-dense regions. The velocity field can also be used to trace the cosmic web {\it backwards} throughout time providing one with better insights into the initial conditions of the early universe \citep{Frisch2001}. \cite{Bond1995} showed that the strain field of the fluid is directly related to the filamentary structure of the cosmic web. Thus, the velocity shear tensor is closely related to the density field as it is a measure of the strain of the DM fluid,  represented by the diagonal elements of the shear tensor. The classification algorithm given by \cite{Hoffman2012} uses information regarding the strain of the fluid, given by the shear tensor, to classify the cosmic web according to the number of eigenvalues of the symmetric shear tensor greater than a given threshold value. Classifying the cosmic web allows one to explore how different web types affect halo properties \citep{Hahn2007a, Hahn2007, Aragon-Calvo2007, Aragon-Calvo2010, Libeskind2012a, Cautun2012, Libeskind2012b, Cautun2014a} and the galaxies within the halos \citep{Nuza2014, Metuki2014}. 

The aim of this article is to compare existing cosmic web classification methods with a new approach using a smoothed particle dynamics (SPH) formalism borrowed from the SPH simulation technique that incorporates information from locally defined neighbours to construct the velocity shear tensor. The approach can be applied to halos directly and produces statistics consistent with current literature \citep{Hoffman2012, Cautun2012,Libeskind2012a, Libeskind2012b, Metuki2014, Nuza2014}. The method allows one to easily incorporate the Hubble flow, which is difficult to account for in grid based approaches. Thus, by accounting for the expansion between particles, it is shown that a natural explanation can be given for the threshold value adopted in grid based approaches.

\begin{figure*}
  \includegraphics[width=1.9\columnwidth]{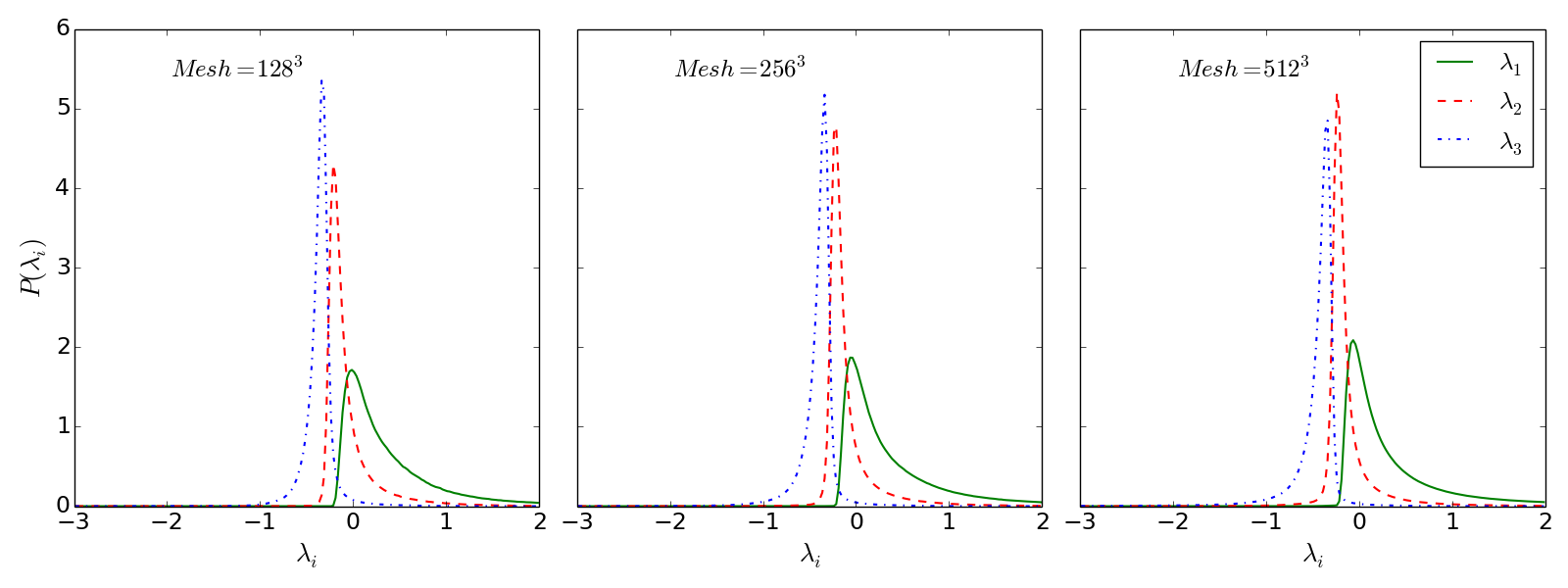} \\
  \includegraphics[width=1.9\columnwidth]{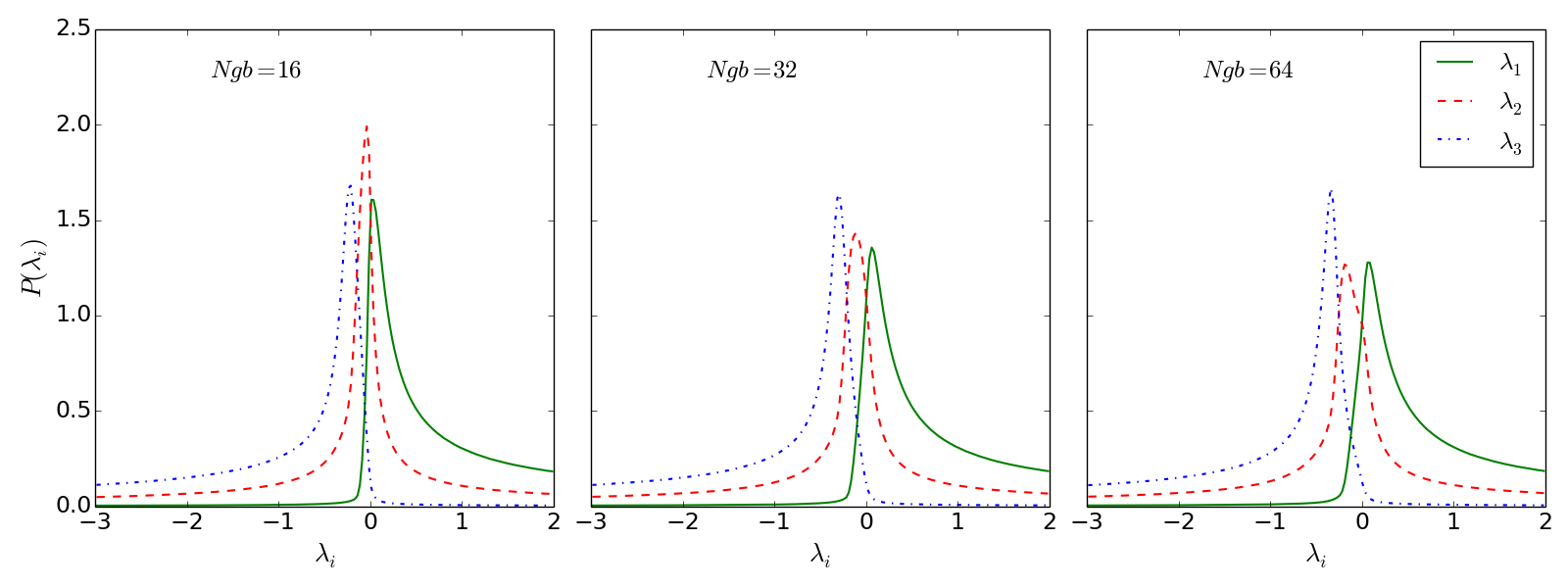} \\
  \includegraphics[width=1.9\columnwidth]{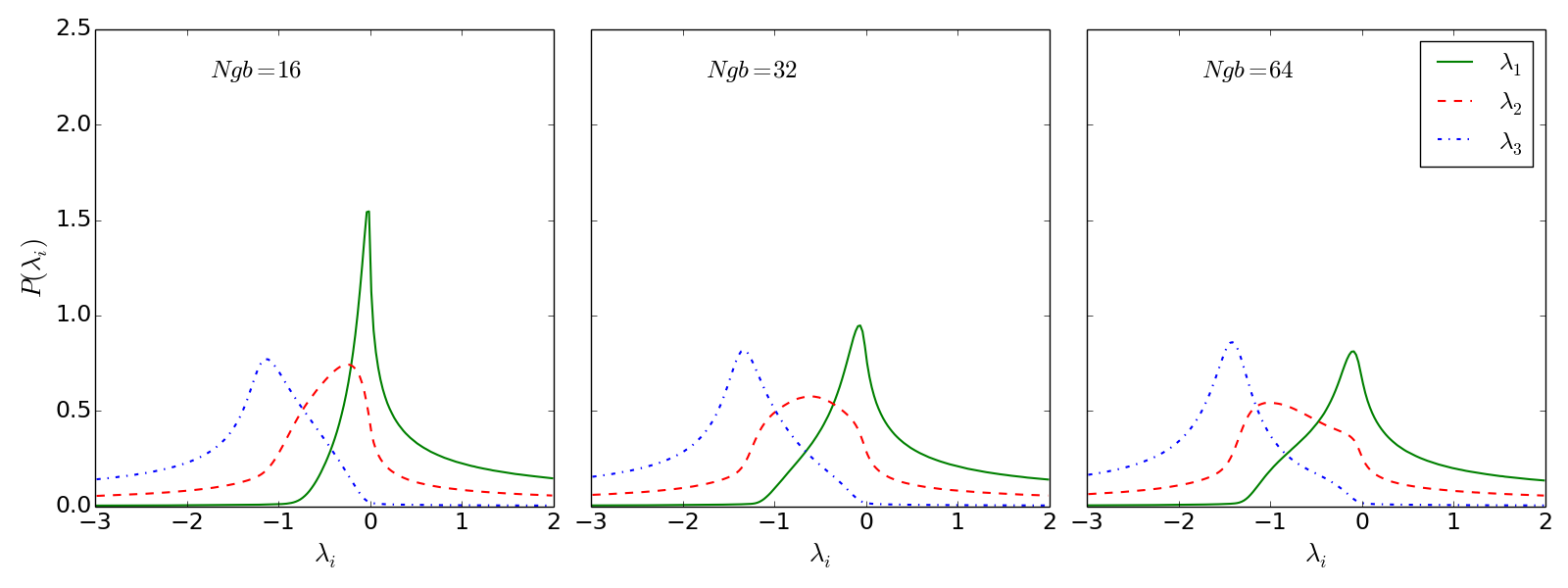} \\
  \caption{\textbf{Top}: The PDFs for eigenvalues using varying grid sizes: $128^3$ (Left), $256^3$ (Middle) and $512^3$ (right). $\lambda_1$ is given by the green solid line, $\lambda_2$ is given by red dashed line and $\lambda_3$ is given by the blue dot-dashed line.  \textbf{Middle}: The PDFs for Lagrangian based eigenvalues using a varying number of neighbours: $16$ (Left), $32$ (Middle) and $64$ (right) with the same line properties as above. \textbf{Bottom}: Lagrangian based PDFs for the $1024^3$ particles incorporating the Hubble flow into the velocities before the shear is determined, using the same neighbour numbers and line properties as above. Distributions are approximately log-normal; a finer mesh is attributed with shorter smoothing lengths or fewer neighbours (neglecting the Hubble flow).}
  \label{eigen_dist}
\end{figure*}
\begin{figure*}
  \includegraphics[width=1.9\columnwidth]{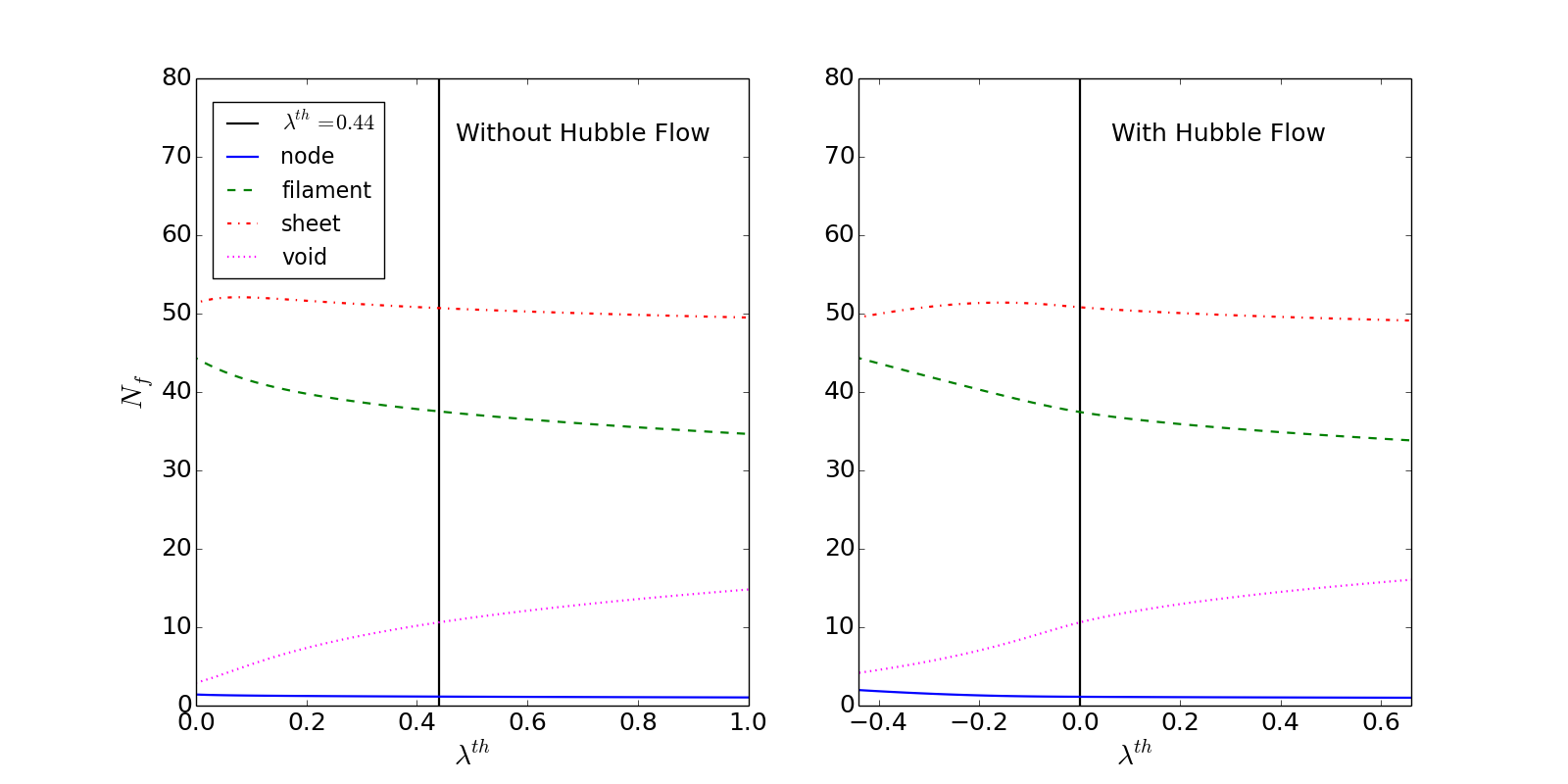} 
  \includegraphics[width=1.9\columnwidth]{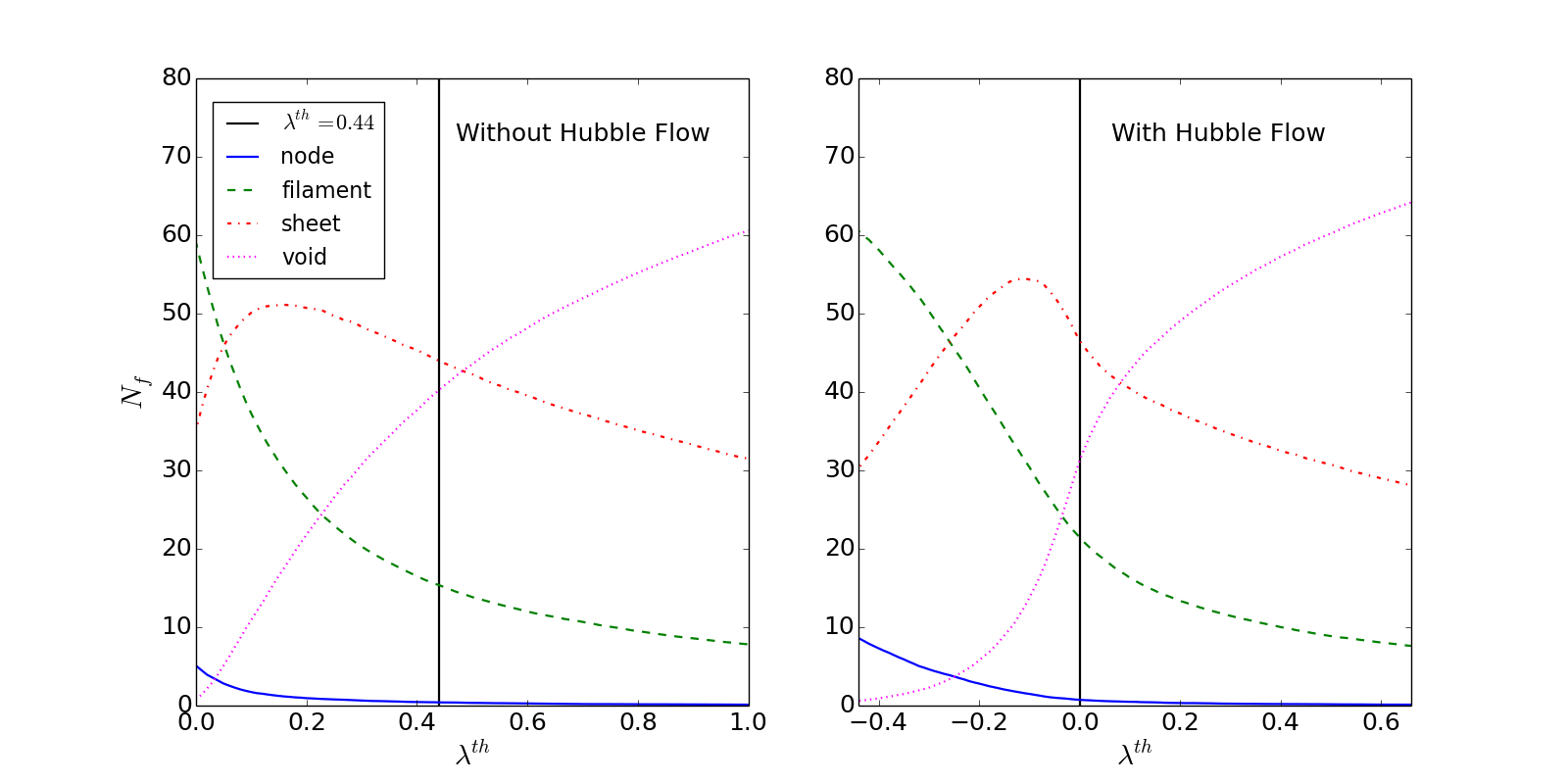} 
  \caption{Changing $\lambda^{th}$ and its effect on the $N_f$ of classified web types for the DM particles (\textbf{top}) and halos (\textbf{bottom}) without (left) and with (right) Hubble flow. Threshold values are given by the black vertical lines at $\lambda^{th} = 0.44$ and $\lambda^{th} = 0.0$ excluding and including the Hubble flow respectively. Nodes are shown by the solid blue line, filaments are dashed green, sheets dot-dashed red and voids dotted magenta. Number fractions are very similar at any threshold value both with and without Hubble flow when shifting the Hubble flow threshold value by $\sim -0.44$.}
  \label{delta_lam}
\end{figure*}
\section{The Simulation}
The following analysis is based on a N-body simulation employing the GADGET-2 simulation code \citep{Springel2005} with a cube length of $64h^{-1}Mpc$, a total of $1024^3$ particles and a particle mass $m_p =  2.1 \times 10^7 h^{-1}M_{\odot} $. Each particle is comprised entirely of  dark matter and the collisional component is ignored. A $\Lambda$CDM cosmological model with Planck 2013 parameters was used such that at z=0 the total matter density $\Omega_{m} = 0.31$ , cosmological constant $\Omega_{\Lambda} = 0.69$, baryon density $\Omega_{B} = 0.049$, Hubble constant $H_0 = 100 \, h\,\, km \,s^{-1}\, Mpc^{-1}$ and Hubble parameter h = 0.678. The primordial power spectrum index, n=0.961 with a normalization value $\sigma_{8} = 0.826$. Classifications were performed at a redshift of 0.
\section{Method}
This section revises the cosmic web classification scheme based on the velocity shear tensor given by \cite{Hoffman2012} and discusses the implementations for the Eulerian and the Lagrangian methods using the full set of simulation particles. Subsequently, we demonstrate that the Lagrangian method can be easily applied to the halo sample, forgoing the need to interpolate the halo classifications from the grid cells. The section concludes with a discussion on the implementation of the Hubble flow which effects a shift of the eigenvalues and, by extension the threshold value needed, permitting a more natural interpretation of the eigenvalues of the velocity shear tensor.

\subsection{Web Classification Using The Velocity Shear Tensor}
The rate of deformation of a fluid is a result of velocity gradients. In 3-dimensional cartesian coordinates, the deformation tensor can be expressed as $\partial v_{\alpha} / \partial r_{\beta}$ for $\alpha, \beta = x, y, z$. One is able to decompose the rank 2 deformation tensor into its shear (symmetric) and  vorticity (anti-symmetric) components respectively to obtain Eq. ~\ref{symmetry}. 
\begin{equation}
  \frac{\partial v_{\alpha}}{\partial r_{\beta}} =  \frac{1}{2}(\frac{\partial v_{\alpha}}{\partial r_{\beta}} + \frac{\partial v_{\beta}}{\partial r_{\alpha}}) + \frac{1}{2}(\frac{\partial v_{\alpha}}{\partial r_{\beta}} - \frac{\partial v_{\beta}}{\partial r_{\alpha}}) 
  \label{symmetry} 
\end{equation}

The shear tensor is of particular interest as the diagonal elements represent the strain (rate of compression or expansion) along a given eigenvector.  \cite{Hoffman2012} defines the symmetric component of the deformation tensor as the V-web velocity shear tensor, $\varSigma_{\alpha \beta}$, scaled by the Hubble constant as:
\begin{equation}
  \varSigma_{\alpha \beta} = - \frac{1}{2 H_0}\left(\frac{\partial v_{\alpha}}{\partial r_{\beta}} + \frac{\partial v_{\beta}}{\partial r_{\alpha}}\right)
  \label{vweb}
\end{equation}
Exploiting a mathematical property of tensor symmetry, the trace of shear tensor is equal to the sum of the eigenvalues and is proportional to the rate of change of density within the simulation for a comoving observer, since: 
\begin{equation}
  \varSigma^{\alpha}_{\quad\alpha} = \lambda_1+\lambda_2+\lambda_3 =  - \frac{1}{H_0} \partial _{\alpha} v^{\alpha} =  \frac{-(\nabla \cdotp \vec{v})}{H_0} \propto \frac{D\rho}{Dt}
  \label{div}
\end{equation}
where here we have denoted $\partial _{\alpha}  v^{\alpha} = \partial v_{\alpha} / \partial r_{\alpha}$ while employing summation notation. Defining the V-web shear tensor with a negative sign (Eq. ~\ref{vweb} ) and sorting the eigenvalues so that $\lambda_1 \ge \lambda_2 \ge \lambda_3$ allows for a useful interpretation as now the largest eigenvalue corresponds to the eigenvector with the fastest rate of collapse. Classification of the cosmic web into four possible environments for dark matter phase space elements is done by counting the number of eigenvalues above a threshold value $\lambda^{th}$. Historically, $\lambda^{th}$ has been chosen by visual inspection of the cosmic web with a popular value being $\lambda^{th} = 0.44$ \citep{Hoffman2012, Libeskind2012a, Libeskind2012b} however this value may vary with different simulation parameters. Voids, sheets, filaments and nodes correspond to  0, 1, 2 or 3 eigenvalues greater than the threshold value respectively. Observe that when all three eigenvalues are 
greater than $\lambda^{th}$, all 3 axes are collapsing yielding a node classification while voids are classified with zero eigenvalues greater than $\lambda^{th}$ indicative of an expansion along all 3 axes. Filaments exhibit a characteristic collapse along 2 axes with an expansion along the third while sheets collapse along 1 axis and expand along 2 axes.
\subsection{Grid Based Method}
\label{sec:grid}
In order to obtain a benchmark model for the subsequent development of the Lagrangian based approach, we first present a classification based on the traditional grid based method. For that purpose, the velocity and density fields are constructed on $128^3$, $256^3$ and $512^3$ grid sizes using `clouds in cells' (CIC) interpolation over 2 cell lengths. 

Increasing the grid size for a fixed particle number inevitably causes empty cells within the grid, particularly at a redshift of zero. The empty cells have an associated zero velocity and impact the CIC algorithm significantly at large grid sizes. For the $256^3$ grid, we find one empty cell in the entire simulation volume whereas for the $512^3$ grid there are $\sim 2 \times 10^6$ empty cells. This is remedied by assigning a mass averaged velocity to the empty cell using the neighbouring cells contained within a sphere of 3 cell lengths. Empty cells are generally located in under-dense regions where the fluid experiences little or no turbulence, thus, the solution is believed to be a reasonable approximation for laminar like flow.  

To remove numerical artefacts introduced by the CIC interpolation, the fields are smoothed with a gaussian kernel using $\sigma = 1$ cell length. For simplicity, the derivatives are evaluated in k-space after employing a Fast Fourier Transform of the velocity field. Each grid cell then has an associated shear tensor and a set of 3 eigenvalues for classification.
\subsection{Lagrangian Method Applied To Dark Matter Particles}
An alternative to grid-based, Eulerian methods for analysing fluid flows is the Lagrangian SPH approach first pioneered by \cite{Lucy1977}, \cite{Gringold1977} whereby the continuous fluid is divided into a set of N discrete fluid elements that are followed throughout the simulation while under the laws of hydrodynamics. SPH methods have long been used for a wide range of astrophysical problems and, due to its Lagrangian nature, the numerical solutions do not suffer from advection errors, possesses good conservation properties and over-densities can be dynamically resolved, see \cite{Hernquist1989, Valdarnini2012, Springel2010} and references within. By analogy one may use a Lagrangian, grid-free, SPH-kernel based method instead of a grid-based, Eulerian method for computing the gradients of the velocity field. In the following, we will address this method as the smoothed particle dynamics (SPD) approach which uses SPH-kernels exclusively for the computation of 
dynamical quantities.

The SPH-approach employs a particle representation of the fluid where each particle has a given mass and an associated volume. The physical properties of the fluid at a given position $\mathbf{r}$, which may be the location of a particle itself, is then given as a function of the properties of the neighbouring particles located at $\mathbf{r'}$ within the smoothing length, h. The smoothing length is an adaptive quantity in that it is commonly chosen to be distance to the {\it n}th neighbour. The neighbouring particles' properties are weighted by a kernel function $W(r, h)$ with $r = |\mathbf{r} - \mathbf{r'}|$. Any smoothed interpolated field $F_s(\mathbf{r})$ of some field $F(\mathbf{r})$ in x dimensions is defined as  
\begin{equation}
  \label{smooth}
  F_s(\mathbf{r}) = \int F(\mathbf{r'}) W(|\mathbf{r}-\mathbf{r'}|, h) d^x\mathbf{r'}\ , 
\end{equation}
where $ W(\mathbf{r}-\mathbf{r'}, h) \rightarrow \delta(\mathbf{r}-\mathbf{r'})$ in the limit as $h \rightarrow 0$. This is essentially a convolution of the neighbouring particle properties with the kernel. 

In this work we adopt the cubic spline given by \citep{Springel2010} with $W(r, h) =  w(q)/h^d$, d is the dimension and $q = r/h$. In 3 dimensions the kernel is given by 
\begin{equation}
  w(q)= \frac{8}{\pi h^3} 
  \begin{cases}
    1-6q^2+6q^3,\qquad  & 0 \le q \le \frac{1}{2} \\
    2(1-q)^3, \qquad  & \frac{1}{2} < q \le 1\\
    0, \qquad & \text{ Otherwise } 
  \end{cases}
  \label{kernel}
\end{equation}
with the kernel dropping to zero at $r = h$ . The kernel has the important property of being normalized to 1. 

Each particle has an associated mass $m_i$ and one can approximate the volume of the kernel occupied by the particle as  $d^3\mathbf{r} = m_i / \rho_i$. Discretising Eq.~\ref{smooth} and using the 'gather approach', one may approximate the smoothed field with j neighbours as
\begin{equation}
  \label{discrete_eq}
  F_s(\mathbf{r}_i) \simeq \sum_{j=1} ^N\frac{m_j}{\rho_j} F_j W(|\mathbf{r}_i-\mathbf{r}_j|, h_i)
\end{equation}

Thus, for the density field,  $F(\mathbf{r}) = \rho(\mathbf{r}) = \rho_i$ and $F(\mathbf{r'}) = \rho(\mathbf{r'}) = \rho_j$ so that the SPH density for the $i^{th}$ particle is given by  
\begin{equation}
  \rho_i \simeq \sum_{j=1} ^N m_j W(|\mathbf{r}_i-\mathbf{r}_j|, h_i)\ . 
\end{equation}

To obtain an expression for the smoothed velocity field, substitute $F(\mathbf{r}) = \mathbf{v_i}$ into Eq. ~\ref{discrete_eq} for the $i^{th}$ particle to obtain
\begin{equation}
  \mathbf{v}_{s, i} \simeq \sum_{j=1} ^N \frac{m_j}{\rho_j}\mathbf{v}_j W(|\mathbf{r}_i-\mathbf{r}_j|, h_i)
  \label{sphvel}
\end{equation}

Continuing with our analysis of cosmic web classification, it is noted in Eq. ~\ref{vweb} that the shear tensor can be calculated using the gradient of the velocity field. The SPD approach exploits a unique property in calculating the velocity gradients; the spatial derivative of a smoothed field results in the derivative only being applied to the kernel, shown in Eq. ~\ref{grad}, for which analytical solutions exist. 
\begin{equation}
  \label{grad}
  \frac{\partial F_s(\mathbf{r}_i)}{\partial \mathbf{r}_i} \simeq \sum_{j=1} ^N \frac{m_j}{\rho_j}F(\mathbf{r}_j) \frac{\partial}{\partial \mathbf{r}_i}W(|\mathbf{r}_i-\mathbf{r}_j|, h_i) 
\end{equation}
where $\partial / \partial \mathbf{r} = \nabla$. For a constant field Eq. ~\ref{grad} is generally non-zero, however using the fact that $\rho (\nabla \mathbf{v}) = \nabla (\rho \mathbf{v}) + (\nabla \rho) \mathbf{v}$ one may obtain Eq. ~\ref{grad_zero} for the velocity gradient which evaluates to zero for constant velocity fields.
\begin{equation}
  \label{grad_zero}
  \frac{\partial \mathbf{v}_{s, i}}{\partial \mathbf{r}_i} \simeq \frac{1}{\rho_i}\sum_{j=1} ^N m_j(\mathbf{v}_i - \mathbf{v}_j)\frac{\partial}{\partial \mathbf{r}_i} W(|\mathbf{r}_i-\mathbf{r}_j|, h_i)
\end{equation}

With the SPD techniques the velocity shear tensor can be defined in terms of neighbouring particle quantities and kernel derivatives. This Lagrangian representation allows one to resolve and classify individual DM particles based on the properties of its neighbouring particles. More generally, it can be applied to any point set, for instance, below we will use DM halos as the basic point set. 

The smoothing length for the $i^{th}$ particle, $h_i$, is determined as the distance to the furthest of the $n$ nearest neighbours (therefore $q \le 1$ in Eq.~\ref{kernel}). The smoothing length for a fixed neighbour number makes this quantity adaptive in the sense that high density environments will generate short smoothing lengths and low density regions will cause long smoothing lengths. An adaptive smoothing length reduces the sampling error for regions of varying density. The variation of nearest neighbour number, $n$, allows one to explore the effect of the smoothing length on the classification of the cosmic web, akin to variable grid sizes (Section ~\ref{sec:grid}).

Using a 3 dimensional cartesian coordinate system such that $\alpha, \beta$ represent coordinates $(x, y, z)$  and $i, j$ as reference to particles to avoid confusion, the $\alpha$ component of the velocity gradient in the $\beta$ direction for the $i^{th}$ particle can then be expressed as:  
\begin{equation}
  \frac{\partial v^{\alpha}_{s,i}}{\partial {\beta}} \simeq \frac{1}{\rho_i}\sum_{j=1} ^N m_j(\mathbf{v}^{\alpha}_i - \mathbf{v}^{\alpha}_j)\frac{\partial}{\partial \beta} W(|\mathbf{r}_i-\mathbf{r}_j|, h_i)\ . 
  \label{grad_component}
\end{equation}

A k-d tree \citep{Kennel2004} is used to find the nearest neighbours for each particle and the shear tensor is constructed using 16, 32 and 64 neighbours as in Eq. ~\ref{vweb} by taking the spatial derivatives of the relevant velocity components defined in terms of the SPD quantities and the spatial derivatives of the cubic spline kernel interpolant in Eq. ~\ref{kernel} (see Appendix ~\ref{derivation}).  
\subsection{Lagrangian Method Applied To Dark Matter Halos}
To locate individual halos the Rockstar halo finder \citep{Behroozi2011} is employed. Plugging in halo positions and velocities into Eq. ~\ref{grad_component} allows one to construct the shear tensor based on the halo sample. In the same manner as for simulation particles, each halo can be classified using the velocity shear tensor \citep{Hoffman2012}. A $50\, km\, s^{-1}$ maximum circular velocity $v_{max}$ cut-off is employed for the for halos. Using the Velocity-Luminosity relation, this corresponds to a r-band cut-off of $\sim$ -13. This limit may neglect some faint dwarf galaxies however the majority of spirals and early type galaxies are well within this range \citep{Trujillo-Gomez2011}. Locating and classifying halos naturally skews the sampling to the densest web types as halos are a biased tracer for the mass of the cosmic web. Classifying individual halos using the halo-based Lagrangian approach is computationally very cheap as halo catalogues are small relative to particle numbers. Finally, it can 
be anticipated that a halo based classification approach can, with some modifications, be applied to observed galaxy samples.    

\subsection{Hubble Flow}
An aspect intrinsically difficult to accommodate in the grid based approach is the inclusion of the Hubble flow. Cosmological N-body simulations utilise comoving coordinates thus the velocities of the simulation particles only reflect peculiar motion. As a consequence, virialised halos which are expected to show a mean radial velocity equal to zero always show a mean inward motion proportional to the radius. In other words, virialised halos shrink if only peculiar velocities are considered. This spurious effect can be easily corrected for by adding the radially outwards pointing Hubble flow, $\propto H d$, with respect to an arbitrary centre (naturally the position of the central object).

The local nature of the Lagrangian approach effortlessly permits the inclusion of the Hubble flow between a given particle and its neighbours. Incorporating the Hubble flow naturally counteracts the artificial contraction between particles which may explain why $\lambda^{th}$ has to be set to some arbitrary value greater than zero when failing to account for the radially outward flow.  Doing so, using $\lambda^{th} = 0$, leads to very similar results when compared to classification without the inclusion of the Hubble flow and setting $\lambda^{th} > 0$. This allows for a straight forward interpretation of the eigenvalues as now negative eigenvalues correspond to expansion and positive eigenvalues indicate contraction along a specific eigenvector. 

\begin{figure}
  \begin{center}
    \includegraphics[width=0.84\columnwidth]{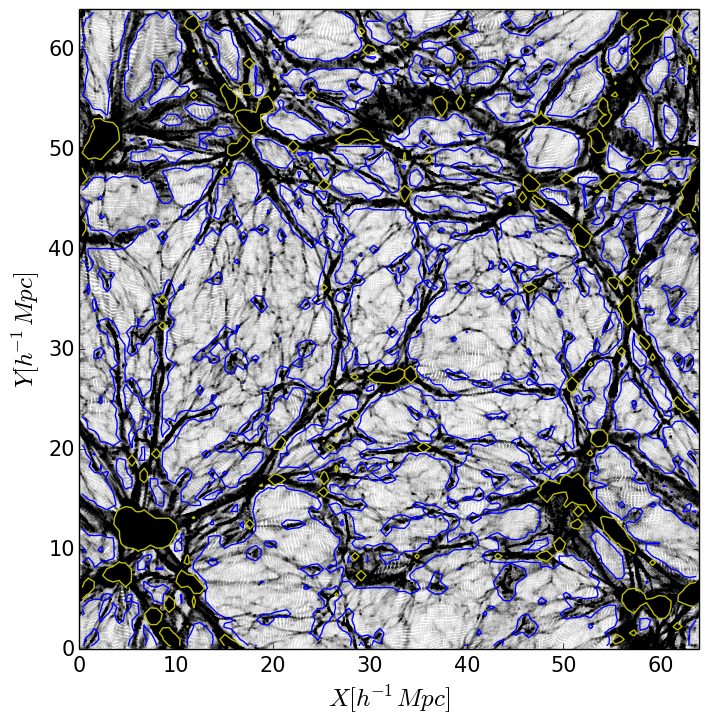}
    \includegraphics[width=0.84\columnwidth]{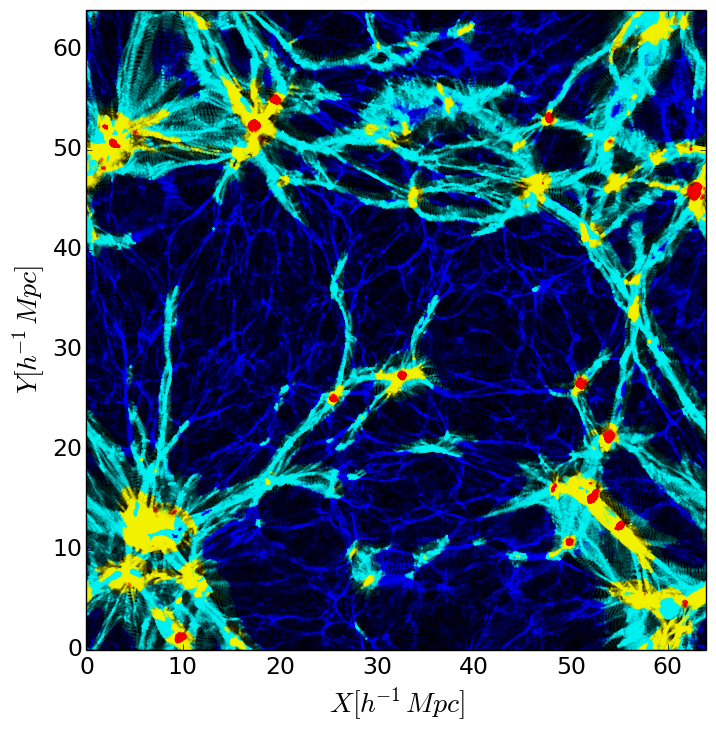}
    \includegraphics[width=0.84\columnwidth]{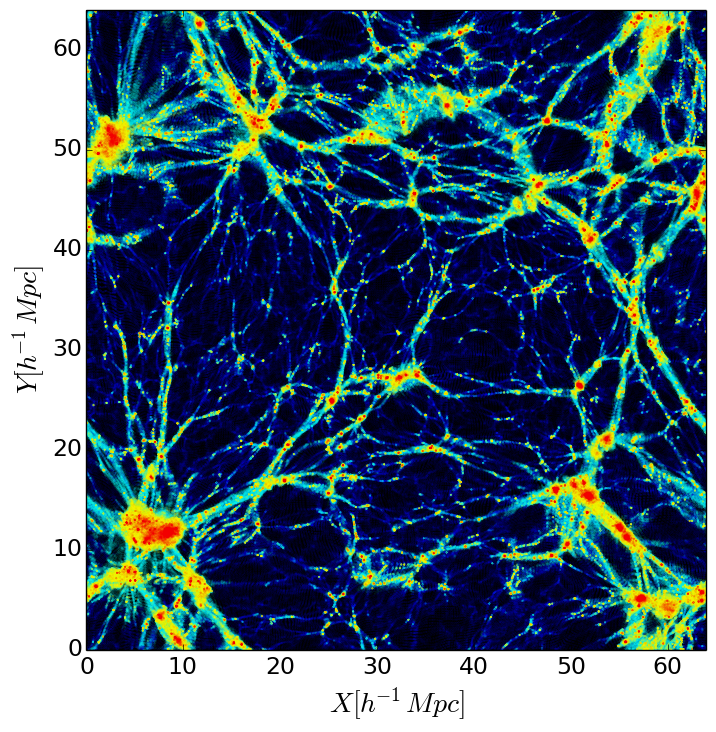}
  \end{center}
  \caption{A slice $0.5h^{-1}Mpc$ thick showing the DM distribution (top), contours indicate regions containing the upper $95^{th}$ (yellow) and upper $68^{th}$ percentile (blue) of the density distribution. The Mesh-based classification using $\lambda^{th} = 0.44$ is shown in the middle sub-figure and the Lagrangian classifier including Hubble flow using $\lambda^{th}=0.0$ is shown in the bottom sub-figure. Voids, sheets, filaments and nodes are represented by the blue, cyan, yellow and red dots respectively.}
  \label{slices}
\end{figure}
\begin{figure*}
  \includegraphics[width=\hsize]{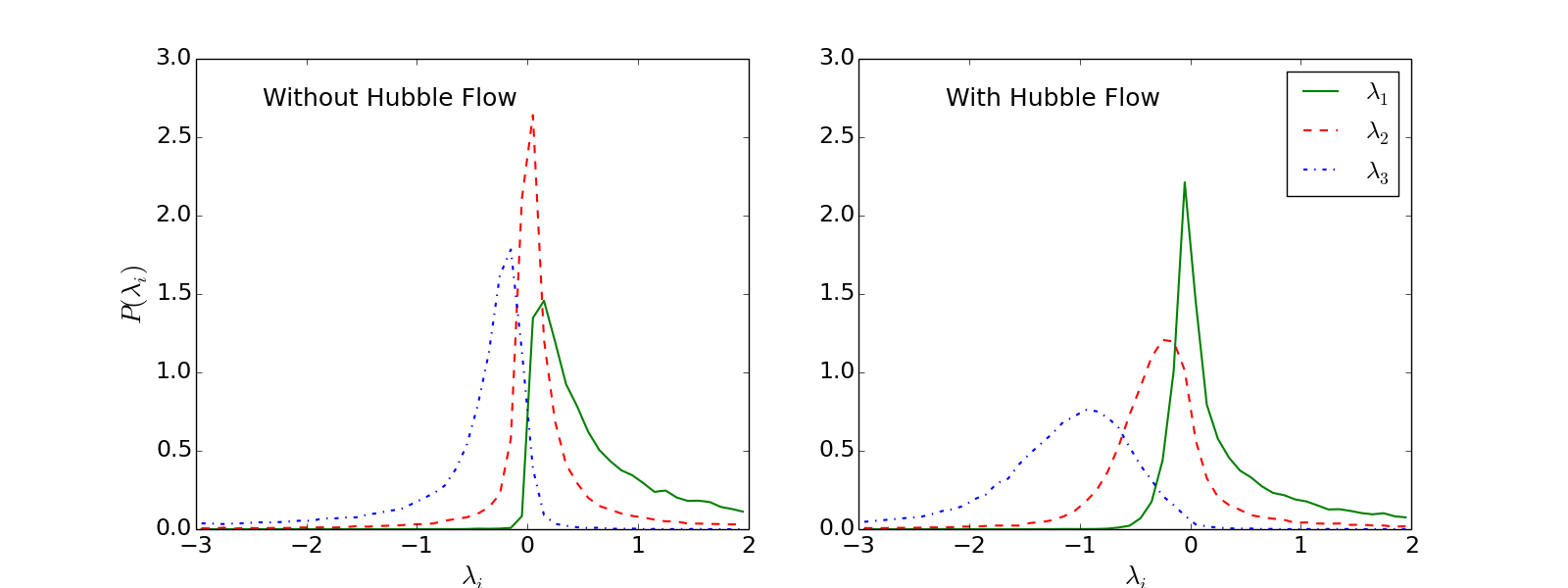} 
  \caption{\textbf{Left}: The normalised PDFs for halo eigenvalues using 32 neighbours without Hubble flow. Colour codes and line styles are defined as in Fig. ~\ref{eigen_dist}. \textbf{Right}: PDFs for halo eigenvalues taking into account the Hubble flow for 32 neighbours. Including the Hubble flow reverses the statistical properties previously observed, a more varied strength of expansion (collapse) is seen along the respective eigenvectors moving from $\lambda_1 \rightarrow \lambda_3$.}
  \label{halo_eigen_dist}
\end{figure*}
\section{Results}
In this section we fist discus the results obtained with the grid based method. This serves as benchmark model which is to be compared with the literature on the one side and with the results based on the SPD method on the other side. Subsequently,  the SPD method is applied to the simulation particle distribution and the halo sample. Finally, the results effected by the Hubble flow correction are discussed.
\subsection{Grid Based Method}
The normalised distribution of eigenvalues for the grid based velocity shear tensor can be found in Fig. ~\ref{eigen_dist} (top), using various grid sizes. The distributions are consistent with the results found in \cite{Libeskind2012b}, following an approximate log-normal distribution. PDFs for the $128^3$ and $256^3$ grids display the characteristic peaks (widths) increasing (decreasing) from $\lambda_1 \rightarrow \lambda_3$. Wider widths in $\lambda_1$ relative to $\lambda_3$ suggest there is more variation in the strength of expansion (collapse) along the corresponding eigenvector. 

A finer grid increases the proportion of $\lambda_3$ eigenvalues found in the lower limit of the range, evident in the decreasing peaks. Distribution peaks for $\lambda_2$ steadily rise for a finer grid, the PDFs for $\lambda_1$ are relatively unchanged. The effects of constraining or relaxing expansion (collapse) through varying grid size is purely numerical.

Volume filling fractions ($V_{ff}$) and mass fractions ($M_f$) for the various grid sizes can be found in Table ~\ref{vmff}. The fractions are in agreement with current literature \citep{Hoffman2012, Metuki2014} and show voids occupy the largest volume of the cosmic web followed by sheets, filaments and nodes. Filaments and sheets are shown to be the most massive web types with nodes and voids making up the balance. 
\begin{table}
  \centering
  \caption{Volume and Mass Filling Fractions: Mesh Based Method Using $\lambda = 0.44$}
  \label{vmff}
  \begin{tabular}{lccccr}
    \hline
    &   Mesh    & Voids    & Sheets    & Filaments  & Nodes \\
    \hline
    $V_{ff}$ & $128^3$   & 67.80\%  & 26.27\%   & 5.40\%     & 0.53\%  \\ 
    & $256^3$   & 67.46\%  & 27.28\%   & 4.87\%     & 0.39\%  \\
    & $512^3$   & 68.93\%  & 26.81\%   & 4.02\%     & 0.24\%   \\
    \hline
    $M_f$ & $128^3$   & 17.88\%  & 31.96\%   & 34.28\%    & 15.88\%  \\ 
    & $256^3$   & 13.80\%  & 36.28\%   & 38.18\%    & 11.74\%    \\
    & $512^3$   & 12.47\%  & 40.63\%   & 38.73\%    & 8.17\%    \\
    \hline
  \end{tabular}  
\end{table}

A finer grid results in lower mass fractions for voids and nodes. This may be explained by the fact that a higher spatial resolution tends to ``de-homogenise'' the eigenvalues, i.e. if all three eigenvalues are above (below) the threshold for a cell at low resolution then the fine structure within the cell resolved by the finer grid causes one or two eigenvalues to shift below (above) the threshold within the new cells transforming voids or nodes into sheets and filaments. This would naturally account for the behaviour of the mass fractions of sheets and filaments which are observed to increase with grid size. The volume filling fractions are relatively consistent for all grid sizes. We use the results based on the $256^3$ grid as the benchmark. 
\subsection{Dark Matter Particles: Lagrangian Method}
Fig. ~\ref{eigen_dist} (middle) shows the distribution of eigenvalues for the Lagrangian approach for all $1024^3$ DM particles neglecting the Hubble flow. The normalised distributions obey similar statistical properties as the PDFs for the grid based method with a log-normal representation and the same eigenvalue width dependency. One observes a similarity between the distributions of finer grids and smaller neighbour numbers; a low neighbour number requirement enforces shorter smoothing lengths, a smoothing scale that is comparable with a very fine grid. Conversely, a coarser grid is a result of a larger smoothing scale, or more neighbours.

Comparing the grid-based eigenvalues with the Lagrangian based eigenvalues excluding Hubble flow, one notes that the peaks for the SPD formulated eigenvalues are shifted slightly towards larger values when compared to their grid based counterparts for any neighbour number or grid size. The SPD-based PDFs tend to have more extended tails with a larger proportion of eigenvalues with absolute values $|\lambda_i|\gg 1$.

SPD results based on 16 neighbours need to be treated with caution due to the low number statistics. Even if the overall trends are similar to the grid based approach, we mainly use the 16 neighbour PDFs to explore the impact of small neighbour numbers on the eigenvalue distribution. The 16 nearest neighbour results are not discussed further. 
      
Using 64 neighbours leads to a spurious behaviour for the PDFs, which is most obvious for the $\lambda_2$ eigenvalue distribution and is evident by a bump around zero. The effect is even more pronounced in Fig. ~\ref{eigen_dist} (bottom right) using 64 neighbours incorporating the Hubble flow. This is believed to be a non-physical artefact as a result of known integral interpolant errors, $\mathcal{O}(h^2)$, associated with larger smoothing lengths due to larger neighbour numbers \citep{Monaghan2005}. 

The Lagrangian method, without Hubble flow, using 32 neighbours produces eigenvalue distributions and mass fraction statistics, seen in Table ~\ref{vmff_sph}, that best match current literature \citep{Hoffman2012, Cautun2012, Libeskind2012b, Metuki2014}. PDFs for 32 neighbours display the properties of decreasing peaks and increasing widths for $\lambda_3 \rightarrow \lambda_1$ which can also be observed in the benchmark grid based PDFs. In addition to this, the choice of ~32 particles is a standard number of neighbours in SPH simulations. For this reason, the SPD halo analysis will resume using 32 neighbours as the benchmark.

Fig. ~\ref{eigen_dist} (bottom) shows the normalised eigenvalue PDFs for the Lagrangian method taking into account the Hubble flow. The expansion (collapse) is again more relaxed relative to the grid based approach, exhibiting more pronounced tails and a larger range. Other statistical properties, such as log-normality and decreasing widths still hold, however, these features are far less obvious. One may observe an approximate symmetry using 32 and 64 neighbours whereby $\lambda_3$ is a negative reflection of $\lambda_1$ around $\lambda_i \sim -0.5$. \cite{Libeskind2012b} suggests this relation would imply expansion and collapse are equivalent processes but in opposite directions.        
\begin{table}
  \centering
  \caption{Mass Fractions: SPH Particle Method With \& Without Hubble Flow Comparing $\lambda^{th}=0.44$ \& $\lambda^{th}=0.0$ for 32 neighbours}
  \label{vmff_sph}
  \begin{tabular}{lcccr}
    \hline
    Without Hubble flow   & Voids    & Sheets    & Filaments  & Nodes \\
    \hline
    $\lambda^{th}=0.44$  & 10.62\%  & 50.70\%   & 37.53\%    & 1.15\%  \\
    $\lambda^{th}=0.0$   & 2.92\%   & 51.36\%   & 44.31\%    & 1.42\%  \\ 
    \hline
    \hline
    With Hubble flow   & Voids    & Sheets    & Filaments  & Nodes \\
    \hline
    $\lambda^{th}=0.44$  & 14.77\%  & 49.50\%   & 34.70\%    & 1.04\%  \\
    $\lambda^{th}=0.0$   & 10.61\%  & 50.81\%   & 37.50\%    & 1.13\%  \\ 
    \hline
  \end{tabular}  
\end{table}

The mass fraction for each web type is shown in Table ~\ref{vmff_sph} with and without Hubble flow using $\lambda^{th}=0.44$ and $\lambda^{th}=0.0$. Using $\lambda^{th}=0.44$ without Hubble flow results in mass fractions for voids and filaments that are  similar to those for the grid classification however sheets occupy a larger proportion of the total mass within the cosmic web while nodes are responsible for a very small amount of the web mass. The low node fractions for all SPD classifications may be attributed to the fact that for a given particle, collapse along all 3 axes is quite difficult to achieve. Each particle experiences a range of tidal forces, due to its neighbouring particle distribution, which in turn will mitigate a complete gravitational collapse inward. Mesh-based approaches do not suffer from this problem as particles are enclosed in individual cells and the CIC interpolation smooths out the gravitational tidal forces to a large degree.

Accounting for the Hubble flow naturally introduces an expansion in all directions, increasing void classifications at any given threshold. Using $\lambda^{th}=0.0$ with Hubble flow produces almost identical mass fractions to $\lambda^{th}=0.44$ without Hubble flow, validating the inclusion of the Hubble flow in the classification calculation. This also explains why it is necessary to introduce a threshold value greater than zero in the grid based approach. 

The top panels of Fig. ~\ref{delta_lam} show the number fraction, $N_f$, for web types as a function of the threshold value $\lambda^{th}$. On the left we find the results without Hubble flow. On the right panel, the Hubble flow is added, i.e. each neighbour particle is attributed with an additional radial velocity according to the Hubble law before the shear tensor is computed. The number fraction curves with and without Hubble flow are almost identical, except that inclusion of the Hubble flow requires a shift of the threshold value towards the left by approximately 0.44. This value roughly corresponds to the threshold value employed for the grid approach. Consequently, the curves show very similar number fractions when a threshold values of 0.44 and 0 are used for the data without and with Hubble flow, respectively.

The bottom panels of Fig. ~\ref{delta_lam} show the same curves for the halo sample. Details will be explored further in Section ~\ref{lag_halo}. Here we only remark that the shapes of curves without (left) and with (right) the Hubble flow are very similar if shifted by $\sim -$0.44, exactly as observed for the particle sample. 

Volume filling fractions are not straightforwardly extractable using the Lagrangian method as one does not classify the entire cosmological volume composed of individual cells but rather individual particles, which does not allow for a trivial summation of the volume.   

The top panel of Fig. ~\ref{slices} shows the DM distribution within a slice of $0.5h^{-1}Mpc$ thickness projected along the z axis onto the x-y plane. The filamentary web structure is obvious and the regions of varying density are clearly identifiable. The yellow and blue contours enclose the regions where the density values are above the 95th and 68th percentile respectively. That is, the top 5\% and top 32\% of the density are within these contours. Yellow contours enclose the densest regions of the slice where many particles have collapsed forming very dark patches. Less dense elongated filaments are captured by the blue contours forming the web like structure in the space between the over-dense regions. 

The middle and bottom panels of Fig.~\ref{slices}  display the same slice classifying each particle using a  $256^3$ grid based method (middle) and 32 neighbours Lagrangian method with Hubble flow (bottom), using $\lambda^{th}=0.44$ and $\lambda^{th}=0.0$, respectively. The SPD-based classification scheme (bottom) resolves more details, for example many more small DM particles within the filaments are identified as nodes. The very same particles are classified as filaments or sheets by the grid based approach. This result is due to the adaptive nature of the SPD approach. Despite the larger number of classified nodes depicted here, the mass fraction for nodes derived from the Lagrangian method is lower compared to the grid based approach. 

Comparing top and bottom panels, it becomes clear that the majority of nodes are associated with the highest density regions. These are contained within the upper $95^{th}$ percentile of the density distribution and are enclosed by the yellow contours. Most of the web mass in the filaments and sheets is then captured within the blue contours leaving only the under-dense voids.   
\begin{figure}
  \includegraphics[width=1.0\columnwidth]{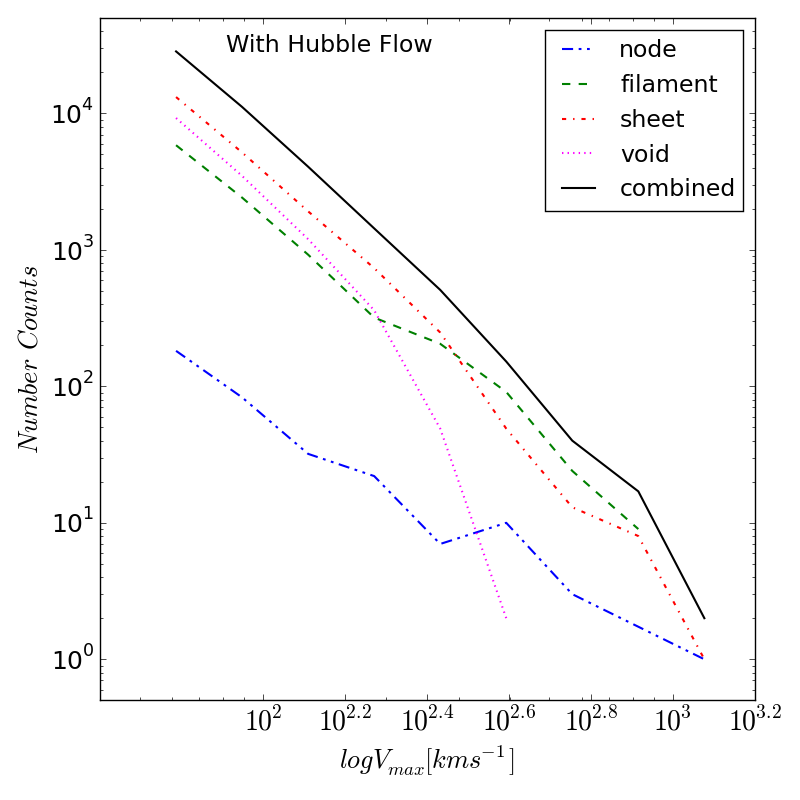}
  \caption{Maximum circular velocity as a function of web type for nodes (blue dashed-dot-dot), filaments (green dashed), sheets (red dashed-dot) and voids (dotted magenta). The computation of the shear tensor was based on the 32 nearest neighbour halos with the Hubble flow added to the halo velocities. For the separation of the eigenvalues (collapsing/expanding) a threshold value of 0.0 was implemented. The velocity function for voids is very steep indicating that there are no high velocity halos in voids. On the other hand, the node velocity function is quite flat allowing for the highest circular velocity halos to be classified as node halos.}
  \label{v_max}
\end{figure}
\begin{figure}
  \includegraphics[width=1.0\columnwidth]{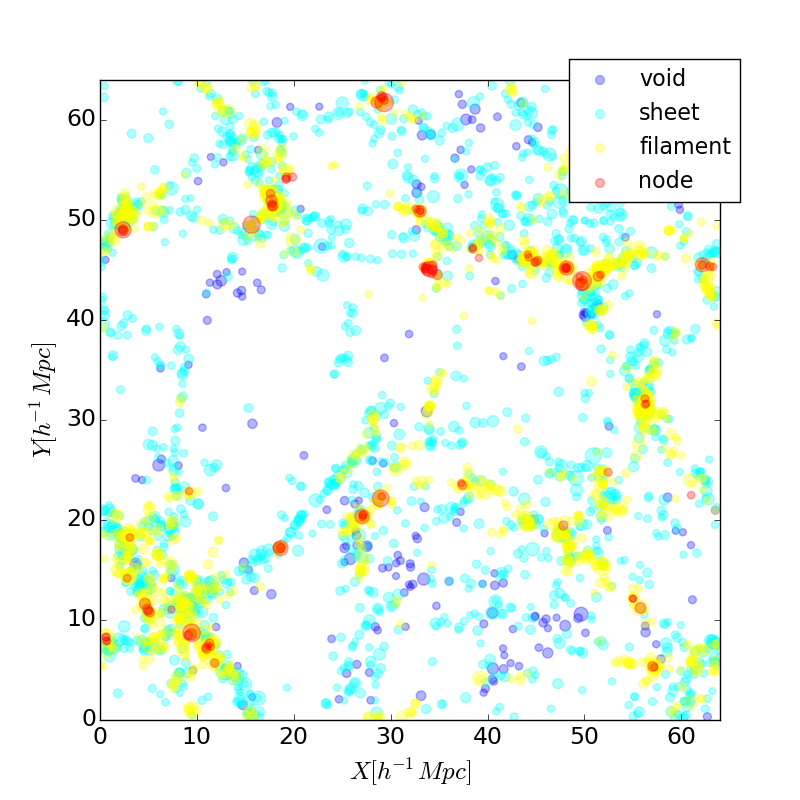}
  \caption{A 5 $h^{-1}Mpc$ slice showing the classified halos using 32 neighbour halos accounting for the Hubble flow. The size of the markers are scaled according to the $v_{max}$ of each halo. Larger halos are generally classified as residing in a denser web type, the smallest halos classified as void halos and the largest as filament and node halos.}
  \label{halo_slice_v_max}
\end{figure}
\subsection{Lagrangian Application to Halos}
\label{lag_halo}
The goal of the cosmic web classification discussed above is to determine the environment that the galaxies inhabit. This allows one to investigate the impact of the environment on galaxy evolution which is crucial to understanding the diversity of galaxy populations. Galaxies are thought to be hosted by dark matter halos. Thus, once the web classification (independent of whether grid or particle based) is completed the various web types have to be assigned to DM halos (i.e., galaxies). This is accomplished with interpolation schemes which allow for the transition between the grid or particle distributions to halo locations. 

Alternatively, the Lagrangian method introduced above can be directly applied to the halo sample, i.e. the velocity shear tensor and its eigenvalues can be computed for an individual halo by utilising the positions and velocities of its nearest neighbour halos. For the subsequent analysis, the velocity shear tensor was computed based on the 32 nearest neighbour halos.

The two bottom panels of Fig. ~\ref{delta_lam} show the fraction of halos belonging to a particular web type as a function of the threshold value $\lambda^{th}$. The bottom left panel of Fig. ~\ref{delta_lam} displays the halo classification without Hubble flow; nodes and filaments are monotonically decreasing with nodes making up a very small percentage as $\lambda^{th}$ increases. Voids are under-represented at low threshold values. Sheet fractions initially rise when moving to a higher threshold value but begin to decrease after $\lambda^{th} \gtrsim 0.15$.

The addition of the Hubble flow in the bottom right panel of Fig. ~\ref{delta_lam} does not drastically change the shape of the number fraction curves for different web types but seems to just shift the curves leftward. Observe the similarity of the curves to the right of the given threshold values (black vertical line) with and without Hubble flow. The threshold values at $\lambda^{th} = 0.44$ without Hubble flow and $\lambda^{th} = 0.0$ with Hubble flow offer very similar number fractions which is further confirmed in Table ~\ref{halo_frac}. As already discussed for the particle based classification scheme, one is compelled to conclude that the determination of $\lambda^{th} = 0.44$ by visual inspection \citep{Hoffman2012} can be understood as an effective quantity accommodating for the Hubble flow lacking in the grid based method. 

The halo eigenvalue PDFs are presented in Fig. ~\ref{halo_eigen_dist} both with and without Hubble flow. The left panel of Fig. ~\ref{halo_eigen_dist} neglects the Hubble flow.  The results are in agreement with both the grid-based as well as the SPD particle based distributions in Fig. ~\ref{eigen_dist} (top \& middle), obeying similar statistical properties. Including the Hubble flow produces the distributions displayed in the right panel of Fig. ~\ref{halo_eigen_dist}). These curves exhibit a reversal of properties, the widths are increasing and peaks are decreasing moving from $\lambda_1 \rightarrow \lambda_3$. Although the method is still SPD based, the wide tails seen in the particle distributions of Fig. ~\ref{eigen_dist} (middle \& bottom) have vanished. This has the effect of constraining the strength of expansion (collapse) in the halo based regime. 

\begin{table}
  \centering
  \caption{SPH Halo Number Fractions ($N_f$) Using 32 Neighbours With \& Without Hubble Flow Comparing $\lambda^{th} = 0.0$ \& $\lambda^{th} = 0.44$  }
  \label{halo_frac}
  \begin{tabular}{lcccr}
    \hline
    Without Hubble flow   & Voids    & Sheets    & Filaments  & Nodes \\
    \hline
    $\lambda^{th}=0.44$  & 40.25\%  & 43.96\%   & 15.37\%    & 0.41\%  \\
    $\lambda^{th}=0.0$   & 0.82\%   & 35.04\%   & 59.03\%    & 5.01\%  \\ 
    \hline
    \hline
    With Hubble flow   & Voids    & Sheets    & Filaments  & Nodes \\
    \hline
    $\lambda^{th}=0.44$  & 58.52\%  & 31.76\%   & 9.52\%    & 0.20\%  \\
    $\lambda^{th}=0.0$   & 31.31\%  & 46.50\%   & 21.44\%    & 0.74\%  \\ 
    \hline
  \end{tabular}  
\end{table}

Maximum circular velocity, $v_{max}$, is a measure of the gravitational potential well for each halo and serves as a proxy for galaxy luminosity. Fig. ~\ref{v_max} shows the maximum circular velocity function for the different web types. Halos classified as nodes are higher density web types and as such make up the majority of high $v_{max}$ halos extending to $\sim 1200 \,km\,s^{-1}$. Under-dense web types such as voids are associated with lower $v_{max}$ values $\lesssim 450 \, km\, s^{-1}$. The majority of halos between 400 $km\,s^{-1}$ and 1000 $km\,s^{-1}$ are classified as filaments which are followed by sheets. We observe a correlation between web type and $v_{max}$ using the SPD approach similar to the mass - web type relation found in \cite{Hahn2007, Cautun2012, Metuki2014}. 

Fig. ~\ref{halo_slice_v_max} shows a 5 $h^{-1}Mpc$ slice of the halos classified using the SPD approach with Hubble flow and $\lambda^{th}=0.0$. The halo size is plotted proportional to the maximum circular velocity. The apparent structure further supports the $v_{max}$ - web type correlation. Voids are generally relatively small circles increasing for sheets, filaments and nodes respectively. Fig.~\ref{halo_slice_v_max} establishes the success of the Lagrangian, halo based, web classification approach. It is an effective way to determine the environment within which halos reside.
\section{Conclusions}
Galaxy formation and evolution is a poorly understood topic and classifying the cosmic web serves as a crucial step in the analysis of the galaxy-environment relationship. Existing methods of cosmic web classification have used Eulerian, grid based, techniques to classify the cosmic web using a visually defined threshold value. 

Here we have proposed a new Lagrangian, SPD-based, method for the classification of the cosmic web. The SPD approach offers an alternative numerical method for classifying the cosmic web that benefits from the SPH implementation used to define particle quantities. The SPD-based classifier draws on the local properties of the underlying discretised density and velocity field to calculate the shearing rate of a fluid in the kernel volume for a given particle or halo. 

The variable smoothing length allows for greater adaptivity, as is evident by the methods' ability to resolve and classify smaller substructure compared to grid based methods. This capability may be an advantage or disadvantage, depending on the problem envisaged.

Our analysis set out to reproduce results obtained with the Eulerian approach as a benchmark. Good agreement between (particle-based) Eulerian and Lagrangian classification schemes is demonstrated. The distribution of eigenvalues for the Eulerian approach and the Lagrangian approach possess similar statistical properties. The mass and number classification fractions in the SPD approach are underestimated for nodes but are otherwise very similar for the remaining web types. The lack of node classifications may be due to the nature of the SPD approach as it captures tidal forces that may prevent a complete collapse along all 3 eigenvectors. 

The Lagrangian representation classifies individual DM particles according to the properties of the nearby particle neighbourhood and allows for an easy inclusion of the Hubble flow between neighbouring particles or halos. The addition of the Hubble flow removes the need for an arbitrary threshold value and sets the threshold value to zero irrespective of simulation parameters.

Classifying the cosmic web using the new SPD scheme is ideally suited to studying galactic properties as galaxies and the halos in which they reside are a function of the external environment and trace the densest regions of the DM phase space. Directly applying the new technique to the halo sample  yields good agreement with halo classifications based on the interpolation of the particle-based Eulerian approach. High maximum circular velocity halos are more likely to be classified as a denser web type. Halos in high density regions, classified as nodes and filaments, exhibit large  $v_{max}$ values whilst under-dense regions such as voids are typically endowed with much smaller $v_{max}$ values. 

\subsection{Acknowledgements}
The authors gratefully acknowledge fruitful discussions with Sean February and the use of computing resources at the Centre for High Performance Computing (CHPC South Africa) as well as the support from the National Astrophysics and Space Science Programme (NASSP).
\bibliographystyle{mnras}

%
\appendix
\section{SPD-based Shear Tensor}
\label{derivation}
The expression for the velocity shear tensor can be derived by rewriting the deformation tensor $\partial v_{\alpha} / \partial r_{\beta}$ as  
\begin{eqnarray}
  \frac{\partial v_{\alpha}}{\partial r_{\beta}} &=& \frac{\partial v_{\alpha}}{\partial r_{\beta}} + \frac{1}{2}\frac{\partial v_{\beta}}{\partial r_{\alpha}} - \frac{1}{2} \frac{\partial v_{\beta}}{\partial r_{\alpha}} \nonumber\\
  &=& \frac{1}{2}(\frac{\partial v_{\alpha}}{\partial r_{\beta}} + \frac{\partial v_{\beta}}{\partial r_{\alpha}}) + \frac{1}{2}(\frac{\partial v_{\alpha}}{\partial r_{\beta}} - \frac{\partial v_{\beta}}{\partial r_{\alpha}}) 
\end{eqnarray}
Where $\alpha, \beta = x, y, z$. The first, symmetric part is used to define the velocity shear tensor scaled by the Hubble factor as  
\begin{equation}
  \varSigma_{\alpha \beta} = - \frac{1}{2 H_0}\left(\frac{\partial v_{\alpha}}{\partial r_{\beta}} + \frac{\partial v_{\beta}}{\partial r_{\alpha}}\right)
\end{equation} 
For a smoothed field using the SPH approach, we have for the $i^{th}$ particle and j neighbours, $\mathbf{r}$ is the position vectors so that
\begin{equation}
  F_s(\mathbf{r}_i) \simeq \sum_{j=1} ^N\frac{m_j}{\rho_j} F_j W(|\mathbf{r}_i-\mathbf{r}_j|, h_i)
\end{equation}
where $h=h_i$ for the `gather' method and $h=h_j$ for the `scatter' method. For the smoothed velocity field  
\begin{equation}
  \mathbf{v}_i \simeq \sum_{j=1} ^N \frac{m_j}{\rho_j}\mathbf{v}_j W(|\mathbf{r}_i-\mathbf{r}_j|, h)\ , 
\end{equation}
the velocity can also expressed as in Eq. ~\ref{grad_zero} using $\nabla (\rho \mathbf{v})_i = \rho_i (\nabla \mathbf{v}_i) + \mathbf{v}_i (\nabla \rho_i)$. Applying the gradients, the derivatives only act on the kernel so that 
\begin{equation}
  \nabla (\rho_i \mathbf{v}) \simeq \sum_{j=1} ^N m_j v_j \nabla W(|\vec{r}_i - \vec{r}_j|, h)
\end{equation}
here we use the fact that $\rho_i = \sum_{j=1} ^N m_j W(|\vec{r}_i - \vec{r}_j|, h)$. Applying the gradient yields $\nabla \rho_i = \sum_{j=1} ^N m_j \nabla W(|\vec{r}_i - \vec{r}_j|, h)$. By rearranging terms in the chain rule equation and substituting the gradient and divergence terms into the equation and solving for $(\nabla \vec{v})_i$ one obtains
\begin{eqnarray}
  \nabla \mathbf{v}_i &\simeq& \frac{1}{\rho_i} \left[ \sum_{j=1} ^N m_j v_j \nabla W(|\mathbf{r}_i - \mathbf{r}_j|, h)
    - v_i \sum_{j=1} ^N m_j \nabla W(|\mathbf{r}_i - \mathbf{r}_j|, h) \right] \nonumber \\ 
  &\approx& \frac{1}{\rho_i}\sum_{j=1} ^N m_j (v_j - v_i) \nabla W(|\mathbf{r}_i - \mathbf{r}_j|, h)  
\end{eqnarray}
The partial derivatives of one component, say the $x$-component of the $i^{th}$ particle $v_{x,i}$, of the velocity is given by : 
\begin{subequations}
  \begin{eqnarray}
    \frac{\partial v_{x,i}}{\partial x_i} &=&   \frac{1}{\rho_{i}}\sum_{j=1} ^N m_j(v_{x,j} - v_{x,i}) 
    \frac{\partial }{\partial x_i} W(|\mathbf{r}_j-\mathbf{r}_i|, h)
  \end{eqnarray}
  \label{x_deriv}
\end{subequations}
Since the kernel is evaluated using $W(|\mathbf{r_i}-\mathbf{r_j}|, h)$, we let $r =  |\mathbf{r}_i - \mathbf{r}_j| = \sqrt{(x_i-x_j)^2 + (y_i-y_j)^2 + (z_i-z_j)^2}$ and $q=r/h$, then for the x component we have the partial derivatives of the kernel function given by the following expressions. The y and z components follow accordingly. First we note that 
\begin{subequations}
  \begin{eqnarray}
    \frac{\partial W(r, h)}{\partial x} = \frac{\partial W(r, h)}{\partial r} \frac{\partial r}{\partial x}
    \label{chain}
  \end{eqnarray}
\end{subequations}
So we may evaluate Eq. ~\ref{x_deriv} using Eq. ~\ref{chain} to obtain 
\begin{subequations}
  \begin{equation}
    \frac{\partial }{\partial x_i} W(|\vec{r}_i-\vec{r}_j|, h)= 
    \begin{cases}
      \frac{8}{\pi h^5} (18 q - 12 ) (x_i-x_j) \qquad  & 0 \le q \le \frac{1}{2} \\
      -\frac{8}{\pi h^5} \frac{6}{q} (1-q)^2 (x_i-x_j) \qquad  & \frac{1}{2} < q \le 1\\
      0 \qquad & \text{ Otherwise } 
    \end{cases}
  \end{equation}
\end{subequations}

%
\bsp	
\label{lastpage}
\end{document}